\documentclass{aastex}
\usepackage{emulateapj5}
\usepackage{epsfig}

\newcommand{\bq}{\begin{equation}}
\newcommand{\eq}{\end{equation}}
\newcommand{\h}{^h}
\newcommand{\m}{^m}
\newcommand{\s}{^s}
\newcommand{\dg}{^\circ}
\newcommand{\p}{$\pm$~}
\newcommand{\am}{'}
\newcommand{\3}{$_3$}
\newcommand{\as}{''}
\newcommand{\simgt}{\lower.5ex\hbox{$\; \buildrel > \over \sim \;$}}
\newcommand{\simlt}{\lower.5ex\hbox{$\; \buildrel < \over \sim \;$}}

\begin{document}
 
\title{NH\3 in the Central 10 pc of the Galaxy I: General Morphology and Kinematic Connections Between the CND and GMCs}
 
\author{
Robeson S. McGary\altaffilmark{1},
Alison L. Coil\altaffilmark{2}, and
Paul T.P. Ho\altaffilmark{1}}

\altaffiltext{1}{Harvard-Smithsonian Center for Astrophysics, 60 Garden Street, Cambridge, MA 02138, \\rmcgary@cfa.harvard.edu, pho@cfa.harvard.edu}
 \altaffiltext{2}{Department of Astronomy, University of California, Berkeley, CA 94720, acoil@astro.berkeley.edu}

\slugcomment{Accepted for publication in ApJ}

\singlespace

\begin{abstract}

New VLA images of NH\3 (1,1), (2,2), and (3,3) emission in the central
10 parsecs of the Galaxy trace filamentary streams of gas, several of
which appear to feed the circumnuclear disk (CND).  The NH\3 images
have a spatial resolution of $16.5''\times14.5''$ and have better
spatial sampling than previous NH\3 observations.  The images show the
``southern streamer,'' ``50 km s$^{-1}$ cloud,'' and new features
including a ``western streamer'', 6 parsecs in length, and a
``northern ridge'' which connects to the CND.  NH\3(3,3) emission is
very similar to 1.2 mm dust emission indicating that NH\3 traces
column density well.  Ratios of the NH\3(2,2) to (1,1) line
intensities give an estimate of the temperature of the gas and
indicate high temperatures close to the nucleus and CND.  The new data
cover a velocity range of 270 km s$^{-1}$, including all velocities
observed in the CND, with a resolution of 9.8 km s$^{-1}$.  Previous
NH\3 observations with higher resolution did not cover the entire
range of velocities seen in the CND.  The large-scale kinematics of
the CND do not resemble a coherent ring or disk. We see evidence for a
high velocity cloud within a projected distance of 50$''$ (2 pc) which
is only seen in NH\3(3,3) and is likely to be hot.  Comparison to 6 cm
continuum emission reveals that much of the NH\3 emission traces the
outer edges of Sgr A East and was probably pushed outward by this
expanding shell.  The connection between the northern ridge (which
appears to be swept up by Sgr A East) and the CND indicates that Sgr A
East and the CND are in close proximity to each other.  Kinematic
evidence for these connections is presented in this paper, while the
full kinematic analysis of the central 10 pc will be presented in
Paper II.
 
\end{abstract}


\section{Introduction}

At a distance of $8.0\pm0.5$ kpc \citep{rei93}, the center of the
Milky Way provides an excellent laboratory for high resolution studies
of galactic nuclei.  Images of radio continuum emission from the
region are dominated by a strong point source, Sgr A*, which is most
likely a $2.6\times10^6$ M$_\odot$ black hole located at the dynamical
center of the Galaxy \citep{mor96,eck97,gen97,ghe98,rei99}.  Sgr A* is
surrounded by a ``mini-spiral'' composed of arcs of ionized gas that
are falling towards and orbiting the black hole \citep{lo83,rob93}.  A
ring of molecular material at a radius of 2 pc from the Galactic
Center surrounds the mini-spiral and is referred to as the
circumnuclear disk (CND)
\citep{gen85,har85,ser86,gat86,gus87,sut90,mar93,mar95}.  The name was
inspired by early observations of HCN and HCO+ which showed signatures
of a single rotating disk with the center swept clear of material
\citep{gus87}.  The western arc of the mini-spiral traces the ionized
inner edge of the CND, while the northern and eastern arms, which
extend towards Sgr A*, may be the result of cloud-cloud collisions in
the CND \citep{lo83,ser85,gen85,gus87,rob93}.

However, with the recent increase in sensitivity and resolution of
instruments, the morphology and kinematics of the gas near the
Galactic Center now appear to be much more complicated than originally
thought.  The CND is composed of many clumps with an average density
of $\sim10^5$ cm$^{-3}$.  Wide-field interferometric mosaics of the
region in HCN(1-0) and HCO+(1-0) \citep{wri00} show that the CND has a
distinct outer edge at $\sim45''$ from the nucleus, making it resemble
a ring more than a disk.  \citet{gus87} observe that the northeast and
southwest lobes of the CND appear to have different angles of
inclination.  While the northeastern sections of the CND are best fit
by an inclination of 70-75$\dg$, the southwestern lobe of the CND
seems to be at an inclination of $\le50\dg$.  This observation led
\citet{gus87} to the conclusion that the CND is warped.  The
southeastern side of the CND is weak in many tracers (including HCN
and HCO+) suggesting that this part of the ``ring'' may be missing
altogether.  \citet{jac93} propose that the two bright lobes of the
CND are in fact {\it distinct} clouds on independent orbits around
the nucleus and do not compose a single ring of gas.  \citet{wri00}
find kinematic evidence that the CND is composed of at least three
distinct clouds which are orbiting the black hole.  In addition, the
northern arm of the mini-spiral actually appears to originate outside
the CND and cross through on its way towards the Galactic Center
\citep{wri00}.  The idea that the CND and the mini-spiral are composed
of multiple distinct clouds under the gravitational influence of the
nucleus easily explains the observed ``warping'' as well as the
complex kinematics and clumpy morphology observed in the region
\citep{den93,jac93,wri00}.

Past observations of the central regions of the Galaxy also provide
insight into questions about the origin of the CND and the clouds
which compose it.  If the CND is composed of many distinct clouds,
then it should be possible to trace where the clouds come from and
determine the origin of the CND.  Many attempts have been made to
detect connections between two nearby giant molecular clouds (GMCs)
and the CND.  \citet{oku89}, \citet{ho91}, and \citet{coi99,coi00}
detect a long filamentary ``streamer'' in NH\3 (1,1) and (2,2)
emission that connects the ``20 km s$^{-1}$ cloud'' (M-0.13--0.08,
\citet{gus81}) to the southeastern edge of the CND.  A small velocity
gradient along this ``southern streamer'' as well as heating and
increased line widths as the streamer approaches the Galactic Center
indicate that gas is flowing from the 20 km s$^{-1}$ cloud towards the
circumnuclear region.  This connection has also been observed in
HCN(3-2) \citep{mar95}, $^{13}$CO(2-1) \citep{zyl90} and 1.1 mm dust
\citep{den93}.  An extension of the 20 km s$^{-1}$ cloud towards the
southwest lobe of the CND was also seen in NH\3 (1,1) and (2,2).  Due
to the lack of pointings to the west of the CND, \citet{coi99,coi00}
could not search for a connection between this extension and the CND.
However, the morphological connection between the 20 km s$^{-1}$ cloud
and the southwest lobe of the CND is observed in dust \citep{zyl98}.
Other signs of connections between the GMCs and the CND include a
possible connection between the northeastern edge of the CND and the
``50 km s$^{-1}$ cloud'' (M-0.03--0.07, \citet{gus81}) observed in
HCN(1-0) \citep{ho93}.

We present maps of the central 10 parsecs of the Galaxy in the (1,1),
(2,2), and (3,3) transitions of NH\3.  Unlike previous NH\3
observations, our new data cover the entire velocity range seen in
molecules at the Galactic Center and have improved spatial sampling
which provide a more complete picture of NH\3 in the central 10 pc of
the Galaxy and enable a more reliable comparison of NH\3 emission to
other tracers.  With this data, we can probe the morphology and
kinematics of all of the CND as well as the surrounding molecular
material. Paper I focuses on the general morphology of velocity
integrated NH\3 emission from the region, including comparisons to
HCN(1-0), HCN(3-2), 1.2 mm dust, and 6 cm continuum emission.  New
features to the north and west of the CND are observed, producing a
complete picture of NH\3 (1,1), (2,2), and (3,3) emission in the
central 10 parsecs.  At least three physical connections between the
GMCs and the CND are observed, with kinematic information presented
for these features.  A detailed discussion of kinematics of the entire
central 10 pc, including temperatures, opacities, and cloud masses,
will be presented in a following paper.

\subsection{{\it Observations}}
 
We observed the (1,1), (2,2), and (3,3) rotation inversion transitions
of NH\3, at 23.694495, 23.722633, and 23.870129 GHz, respectively,
with the National Radio Astronomy Observatory's Very Large
Array\footnote{The National Radio Astronomy Observatory is a facility
of the National Science Foundation operated under cooperative
agreement by Associated Universities, Inc.}  (VLA) telescope in 1999
March.  Observations were made in the D-northC array which provides
the shortest baselines and the most circular synthesized beam possible
for this low declination source ($\sim2''$). A five-pointing mosaic
was used to fully sample the central 4$'$ (10 pc) of the Galaxy.  The
full width at half maximum (FWHM) of the primary beam is 125$''$ at 23 GHz
for the 25 m VLA antennas.  To ensure the field was properly sampled,
one pointing was centered on Sgr A*
($\alpha_{2000}=$17$^h$45$^m$40$^s.0$,
$\delta_{2000}=-29^{\circ}00'26''.6$) while the other four pointings
were offset by 1$'$ to the northeast, northwest, southeast, and
southwest.  At least one hour on source was obtained for each pointing
at each frequency.  The data consist of 31 channels with a width of
9.8 km s$^{-1}$ (0.78 MHz) centered on a $v_{LSR}$ of --10.6 km
s$^{-1}$ and covering $\pm150$ km s$^{-1}$.

Coil \& Ho (1999,2000) imaged the circumnuclear region in NH\3 (1,1)
and (2,2) with a total velocity coverage of $\sim130$ km s$^{-1}$ and
a velocity resolution of 5 km s$^{-1}$.  In their mosaics, fields on
the CND were centered on $v_{LSR}=-10.6$ km s$^{-1}$ while fields to
the south and east of the CND were centered on $v_{LSR}=+31.1$ km
s$^{-1}$.  The narrow velocity coverage excluded high velocity
emission which is seen up to \p110 km s$^{-1}$ in the CND. In
addition, \citet{coi99,coi00} focused on the southern streamer and the
50 km s$^{-1}$ cloud and had no pointings to the north or west of the
CND.  The images presented in this paper represent the first time that
the (1,1), (2,2), and (3,3) transitions of NH\3 have been fully
sampled over the central 4$'$ with a velocity coverage that includes
all of the emission from the CND.

\section{{\it Data Reduction}}

Each pointing was calibrated separately using NRAO's Astronomical
Imaging Processing System ($AIPS$).  Anomalously high $uv$ points and
large closure errors were removed. Calibration of the amplitudes and
phases were performed using 1331+305 (3C286) and 1733--130.  Since Sgr
A* is a strong source at 23 GHz and within the primary beam of each
pointing, self-calibration was successfully applied to each of the
fields.  The calibrator 1229+020 was used for bandpass calibration to
correct instrumental phase offsets across the band and to remove the
roll-off in sensitivity at the edge of the band.  Channels 1, 30, and
31 could not be fully corrected and were removed from the dataset
resulting in a velocity coverage of --140 to +130 km s$^{-1}$.  Once
the data were calibrated, line-free channels were used to model the
continuum emission and remove it from the data.

The fully calibrated, continuum-free data cubes were imaged in MIRIAD
\citep{sau95}, which offers superior tools for making mosaics and
analyzing the spectral information.  The five pointings were combined
into a single mosaic in the $uv$ plane.  For the maps presented in
this paper, we used the maximum entropy method (MEM) to deconvolve the
images.  Previous NH\3 mosaics were deconvolved using the CLEAN
algorithm in which the field is assumed to be composed of points
sources.  As a result, negative bowls surround bright extended
features, possibly obscuring faint connections between clouds (see
\citet{coi99,coi00}).  MEM is better suited to handle fields of
extended emission and does not produce these negative bowls.  However,
MEM does not conserve the integrated flux density in the map.  To
check our deconvolution, we deconvolved the data using both CLEAN and
MEM.  The two images agree on every main feature with the only
difference in the low-level emission surrounding bright features.  The
integrated flux densities in both maps are also equal indicating that
MEM did not add substantial flux to the image.  In order to recover
the low-level emission in the map, the mosaic was imaged using a
Gaussian taper applied to the $uv$ data resulting in a synthesized
beam of $\sim15''\times13''$ at a position angle of $\sim0^\circ$ in
all three maps.

\section{{\it Morphology of the Central 10 pc}}

The NH\3 (1,1), (2,2), and (3,3) velocity integrated (270 km s$^{-1}$)
images are shown in Figures \ref{11.fig}, \ref{22.fig}, and
\ref{33.fig}.  In all three images, the approximate location of the
CND \citep{mar93} is represented by two ellipses and the location of
Sgr A* is marked by a cross.  If a straight summation of each channel
map is used to create the velocity integrated image, emission from
narrow-line features, which appear in only a small fraction of the
channels, is diluted by the increased noise in the summed map.  In
order to avoid this problem and present a map showing all of the
emission in the region, each channel map was clipped below $2\sigma$
before the channels were summed.  This procedure was performed in AIPS
using the IFLUX parameter in MOMNT.  The resulting image had features
on the size scale of single pixels as a result of the blanking
process.  To remove these artifacts, the image was convolved with a
gaussian beam with FWHM=$6''\times6''$.  Since the gaussian was
significantly narrower than the image beam, the final beam size for our
images was only slightly increased to $\sim16.5''\times14.5''$.

If the noise in one channel is known, then the noise in the velocity
integrated image (in Jy beam$^{-1}$ km s$^{-1}$) is simply $\sqrt
N\Delta v\sigma_i$, where $N$ is the number of channels in the
summation, $\Delta v$ is the channel width in km s$^{-1}$, and
$\sigma_i$ is the noise in one channel.  Since the maps were clipped
below $2\sigma_i$ in each channel, $N$ varies for each pixel in our
maps.  However, we assume most features show emission in $\frac{1}{4}$
of the channels and adopt $\sigma=\sqrt 7\Delta v\sigma_i$, where
$\Delta v=9.8$km s$^{-1}$ and $\sigma_i$ is measured in the individual
channels.  The resulting noise in our images is $\sigma_{11}=0.28$ Jy
beam$^{-1}$ km s$^{-1}$, $\sigma_{22}=0.30$ Jy beam$^{-1}$ km
s$^{-1}$, and $\sigma_{33}=0.33$ Jy beam$^{-1}$ km s$^{-1}$ for the
NH\3 (1,1), (2,2) and (3,3) velocity integrated images, respectively.
For features with emission in a very narrow velocity range, the
signal-to-noise (S/N) will be {\it underestimated} in our map by a
factor of $\frac{\sqrt 7}{\sqrt{N'}}$ where $N'$ is the true number of
channels in which the feature appeared.  Thus, a feature with emission
in only one channel will have $N'=1$ and the S/N underestimated by a
factor of 2.6.  The S/N of broad velocity features is {\it
overestimated} in our images by a factor of $\frac{\sqrt{N'}}{\sqrt
7}$ and the S/N for a feature with emission in all channels is
overestimated by a factor of 2.  There are few examples of these
extremes in our data and the contours presented here are a reliable
indicator of the significance of the detections.  The additional
complication in the noise analysis is well worth the benefits of being
able to see all of the emission from the Galactic Center without
losing narrow-line features due to the increased noise of a straight
summed map.

For reference, we also present the NH\3 (3,3) channel maps in Figure
\ref{chan.fig}.  No clipping has been performed on these maps and the
noise should be constant at all pixels (ignoring any effect of MEM
deconvolution).  The maps were not convolved by the $6''$ gaussian
beam and have a resolution of $15.5''\times13.2''$.  Narrow-line
features such as the northern ridge are clearly visible.

In addition to the velocity integrated images, a line ratio map of
NH\3 (2,2) to (1,1) emission is shown in Figure \ref{temp.fig}.  Line
ratios of the NH\3 transitions are reliable indicators of gas
temperatures \citep{ho83,wal83,dan88} and can be helpful in understanding
conditions near the Galactic Center.  Only points with NH\3(1,1)
emission $>3\sigma_{11}$ and NH\3(2,2) emission $>3\sigma_{22}$ are
included in the map, and the single contour marks a $3\sigma_{22}$
detection of NH\3(2,2) emission.  Thus, regions {\it inside} this
contour and with {\it no} greyscale were excluded on the basis of low
NH\3(1,1) brightness and may be very hot with high line ratios.

Gas at the location of the CND is seen in all three velocity
integrated images although it is best traced by NH\3 (3,3).  NH\3(3,3)
is expected to be brighter than NH\3(1,1) and (2,2) by a factor of 2
due to a quantum mechanical degeneracy of the states in the ortho
species of NH\3 \citep{ho83}.  The southern half of the CND is much
stronger in NH\3 (2,2) than (1,1) and Figure \ref{temp.fig} shows the
increased line ratio in the region.  Assuming the CND is optically
thin in NH\3 (1,1) and (2,2) (consistent with the results of
\citet{coi99,coi00}), the observed ratio of NH\3 (2,2) to (1,1)
emission of 1.5 implies a rotational temperature of $\sim$50 K and a kinetic
temperature of 60 K in the southeastern half of the CND
\citep{dan88,ho83}.  This is hotter than material farther from the
nucleus.  Emission to the south and east of the CND has a typical line
ratio $\simlt 1$ indicating a rotational temperature of no more than
$\sim30$ K for gas more than 2 pc to the south or east of the Galactic
Center

The NH\3 images contain many long filamentary features which we refer
to as streamers. The ``southern streamer''
\citep{oku89,ho91,coi99,coi00}, labeled in Figure \ref{33.fig}, can be
seen beginning at $17^h45^m40^s, -29^\circ03'00''$ moving to the
northeast to $17^h45^m42^s.5$, $-29^\circ01'40''$ and intersecting the
CND at $17^h45^m41^s, -29^\circ00'55''$.  Two clouds, previously
observed by \citet{ho91}, are located to the east of the southern
streamer and are labeled as ``SE1'' and ``SE2''.  These clouds appear
to connect to the southern streamer in all three images.  The bright,
``y''-shaped cloud in the northeast centered on $17^h45^m49^s,
-28^\circ59'30''$ is the ``50 km s$^{-1}$'' giant molecular cloud
which is being impacted on the west by Sgr A East (see section 4.2).
Emission from SE1, SE2, and the southern streamer connect the 20 km
s$^{-1}$ cloud in the south to the 50 km s$^{-1}$ cloud in the east.
In fact, SE2 is the southern half of the ``molecular ridge'' discussed
by \citet{coi00}.

The detection of SE1 is a good example of the improved quality of our new images
compared to previous NH\3 mosaics.  Although the new data still show
mainly filamentary structures, there are more faint connections
between clouds than in previous images by \citet{coi99,coi00}.  Only
faint emission was seen from SE1 in \citet{coi00}, while we find this
feature to be as strong as SE2.  Our ability to detect SE1
reflects the lack of negative bowls near bright structures when MEM is
used to deconvolve the images, whereas only CLEAN was used by
\citet{coi99,coi00}.  Our improved spatial sampling and velocity
coverage as well as the use of MEM deconvolution has resulted in more
robust NH\3 images.

A new ridge of NH\3(3,3) emission is seen in Figure \ref{33.fig} to
the north of the CND from $17^h45^m47^s$, $-28^\circ58'15''$ to
$17\h45\m44\s$, $-28\dg58\am50\as$ and lies along the northern edge of
Sgr A East.  In the NH\3(1,1) and (3,3) images, this ``northern ridge'' is
connected to the CND at $17\h45\m43\s.5$, $-29\dg59'45''$ by a narrow
streamer .  The northern ridge was not in the field of view of the
\citet{coi99,coi00} mosaic.  Finally, there is a long ``western
streamer'' running approximately north-south from $-29^\circ02'30''$
to $-28^\circ59'45''$. Multiple faint connections between the western
streamer and the CND are seen at $17\h45\m37\s.5$, $-29\dg01\am40\as$,
$17\h45\m36\s.5$, $-29\dg00\am55\as$, 
and $17\h45\m37\s$, $-29\dg00\am00\as$ and are discussed in more
detail in Section 5.

\section{Comparison to Other Tracers}

The large velocity coverage and uniform spatial sampling of our NH\3
images present a unique opportunity to compare NH\3 emission in the
Galactic Center to emission from other tracers.  In the following
sections, NH\3 (3,3) emission is compared to HCN(1-0), HCN(3-2), 1.2
mm dust, and 6 cm continuum emission.

\subsection{{\it Comparison to HCN(1-0)}}

Comparison of NH\3(3,3) to HCN($J=$1-0) has proved difficult with the
two molecules apparently tracing different features at the Galactic
Center \citep{coi99,coi00,wri00}.  However, the images that were
compared were often unevenly sampled or had very different velocity
coverage and resolution.  Figure \ref{hcn.fig} shows the velocity
integrated NH\3 (3,3) image in contours overlaid on a HCN(1-0) mosaic
of the central 12.5 pc by \citet{wri00}.  The NH\3(3,3) image has been
convolved with a Gaussian taper to a resolution of
$16.6''\times14.5''$ and the HCN(1-0) image has a resolution of
$13''\times4''$.  Although the NH\3 mosaic was made from VLA data and
the HCN(1-0) image was made from a 19-pointing BIMA mosaic, these two
images are similar in their velocity and spatial coverage.  The
NH\3(3,3) image is fully sampled out to 4$'$ with a velocity coverage
of --140 km s$^{-1}$ to +130 km s$^{-1}$ while the HCN(1-0) image is
fully sampled out to 5$'$ and has a slightly broader velocity coverage
of $\pm170$ km s$^{-1}$.  The velocity resolution of the HCN(1-0)
image is 5 km s$^{-1}$ whereas the NH\3 resolution is 10 km s$^{-1}$.
The high quality of these two datasets enables the most robust
comparison of emission from NH\3(3,3) and HCN(1-0).

\subsubsection{{\it Comparison of emission from the CND}}
The ring-like structure of the CND is clearly seen as the backwards
``C'' in the center of the HCN(1-0) image.  Both molecules show peaks
in the northeast and southwest of the CND, but the peaks are not in
the same location.  Although the relative intensities differ, there is
emission from both molecules around the entire ring except in the
southeast where there is a conspicuous lack of emission.  A narrow,
north-south gap in the northern part of the CND is observable in
HCN(1-0), and there is a distinct edge in the NH\3(3,3) emission
coincident with the edge of the gap. While the HCN(1-0)
emission appears to have been pulled into a long filament at the gap
($17\h45\m40\s$, $-28\dg59'$ to $-29\dg00'$), the NH\3 does not appear
to be as extended towards the north.  However, the resolution of the
NH\3 maps makes it difficult to trace small scale features which can
be seen in the HCN(1-0) image.  This gap may be the result of the
northern arm crossing through the CND (see Section 4.3).

Kinematic information must be used to determine whether NH\3(3,3) is
tracing the same features of the CND observed in HCN(1-0).  Figure
\ref{specpos} shows the location of 12 spectra near the CND.  The
corresponding spectra are shown in Figure \ref{spec}.  Spectra A - J
are at the same locations as the HCN(1-0) spectra in \citet{wri00};
spectra K and L are located at two bright NH\3 features.  Spectra for
these two features were not shown in \citet{wri00}, but we include
them here for comparison to NH\3.  The NH\3(3,3) spectra are shown in
thick lines with the HCN(1-0) spectra shown in thin lines, scaled down
by a factor of 8.

Spectrum A is centered on Sgr A*.  HCN(1-0) shows significant
absorption from known clouds along the line-of-sight (\citet{wri00}
and references therein).  NH\3(3,3) shows some weak absorption at 0 km
s$^{-1}$, but also some emission near 40 km s$^{-1}$.  Less absorption
along the line-of-sight in NH\3(3,3) is due to the low optical depth
$\tau<1$ of NH\3(3,3) compared to HCN(1-0) which is optically thick
over most of the region \citep{wri00}.

Spectra E and G show features at the same velocity in NH\3(3,3) and
HCN(1-0).  In spectrum E, both NH\3(3,3) and HCN(1-0) show a feature
at 70 km s$^{-1}$, but HCN(1-0) shows an additional feature at --60 km
s$^{-1}$.  In spectrum G, there are HCN(1-0) and NH\3 features at both
+50 and +80 km s$^{-1}$.  Although the dip in the profile looks like
it could be the result of self-absorption, position-velocity cuts in
NH\3(3,3) through the western side of the CND indicate that this
spectrum is the result of the superposition of two clouds along the
line-of-sight.  In the spectra that agree kinematically, the relative
intensities of the lines are often different.  NH\3(3,3) is especially
weak in high velocity gas.  In HCN(1-0), spectrum I is dominated by a
single feature at --110 km s$^{-1}$. This feature is seen in
NH\3(3,3), but it is weak and surrounded by faint features at --20,
20, and 50 km s$^{-1}$.  In spectrum C, there is bright HCN(1-0)
emission at +100 km s$^{-1}$, but there is no significant detection of
this feature in NH\3(3,3).

Several NH\3(3,3) spectra have a peak velocity 10-30 km s$^{-1}$ less
than HCN(1-0).  Spectrum B peaks at 60 km s$^{-1}$ in NH\3(3,3) and 75
km s$^{-1}$ in HCN(1-0).  Spectrum F shows a similar difference and
peaks at 50 km s$^{-1}$ in NH\3(3,3) and 60 km s$^{-1}$ in HCN(1-0).
In spectrum D, the NH\3(3,3) emission is 30 km s$^{-1}$ lower than the
HCN(1-0) emission.  While the HCN(1-0) emission peaks at 70 km
s$^{-1}$, we see the NH\3 peak at $\sim40$ km s$^{-1}$.  In
\citet{wri00}, spectrum B is nearly crossed by position-velocity cut
$e$ (see Figure 13$e$, position 25$''$, \citet{wri00}).  In this
position-velocity diagram, absorption is seen at nearby positions from
60 to 80 km s$^{-1}$.  This absorption may affect spectrum B, causing
the profile of this spectrum to have a sharp cut-off below 80 km
s$^{-1}$.  If the low velocity emission has been absorbed, then the
true distribution of HCN(1-0) emission may actually be centered at a
lower velocity and agree with the peak velocity of 60 km s$^{-1}$ seen
in NH\3(3,3) emission.  Similar indications of absorption are seen in
position velocity cuts of HCN(1-0) emission near Spectra D and F.  In
HCN(1-0), spectrum D appears to be affected by absorption at 60 km
s$^{-1}$ (see Figure 13$c$, position 20$''$, \citet{wri00}).  Spectrum
F has an asymmetric profile in HCN(1-0).  Although this may be the
result of a shock, it could also be attenuated at low velocities by
HCN(1-0) absorption.  There is absorption in HCN(1-0) from --60 to +50
km s$^{-1}$ near the position of spectrum F (see Figure 13$c$,
position --50$''$, \citet{wri00}).

Spectra H and J indicate self-absorption of HCN(1-0).  In
\citet{wri00}, many HCN(1-0) features showed evidence for
self-absorption, especially in the position-velocity diagrams.  In
spectrum H, NH\3(3,3) appears to peak between the HCN(1-0) emission at
+20 and --40 km s$^{-1}$ indicating that the feature at --10 km
s$^{-1}$ may have been self-absorbed in HCN(1-0) leaving only the
wings of the emission.  In spectrum J, the two HCN(1-0) peaks at --40
and --70 km s$^{-1}$ could be the wings of a self-absorbed spectrum.

Some of the brightest NH\3(3,3) emission from the CND is seen in
spectra K and L.  There is little HCN(1-0) emission in K, while in L
the HCN(1-0) is shifted to a higher velocity.  The shift in L is
similar to spectrum B which is near the same location in the CND.  In
NH\3(3,3), both K and L have a very narrow profile in NH\3 with a FWHM
of $\simlt20$ km s$^{-1}$.  These profiles are much narrower than all
other spectra in the CND.

The general agreement of the NH\3(3,3) and HCN(1-0) spectra,
especially when absorption along the line-of-sight and self-absorption
are considered, shows that NH\3(3,3) traces the main features of the
CND.  The complex spectra in both molecules are an indication of the
complicated gas kinematics in the CND.  When the NH\3(3,3) data are
considered, it becomes even more evident that the CND is not a
coherent ring, but rather is composed of many distinct features.  The
difference in relative intensity of NH\3(3,3) and HCN(1-0) may be
explained by opacity or a variety of chemical effects including
evaporation of NH\3 off of dust grains (see \citet{wri00} for a more
detailed discussion).

\subsubsection{{\it Comparison of emission outside the CND}}

It is also important to compare NH\3(3,3) and HCN(1-0) emission
outside the CND.  Overall, the HCN(1-0) appears to have a more diffuse
distribution with faint emission seen over most of the field.  An
exception is the northwest quadrant of the image which has little
emission from either molecule.  This region is also empty in dust (see
Figure \ref{pbcor}).  There is emission from the 50 km s$^{-1}$ cloud
in HCN(1-0) and NH\3 although it is concentrated in different places.
HCN(1-0) peaks to the north of the NH\3 cloud with the strongest
emission at the connection between the 50 km s$^{-1}$ cloud and the
northern ridge at $17\h45\m47\s$, $-28\dg58'30''$.  The most notable
difference in this region of the 50 km s$^{-1}$ cloud is the ``hook''
of HCN(1-0) emission which extends past the edge of the NH\3 emission
to $17\h45\m46\s.5$, $-29\dg00'15''$.  This location is interior to
the edge of Sgr A East in projection and has a velocity of --80 km
s$^{-1}$ in HCN(1-0).

The southern streamer is weak in HCN(1-0).  There is almost no
HCN(1-0) emission at the peak of the NH\3 emission at $17\h45\m43\s$,
$-29\dg01'30''$.  Some HCN(1-0) emission is coincident with SE1 and
SE2, but the features are not as connected as in NH\3.  The southern
third of the western streamer is seen in both HCN(1-0) and NH\3.  This
feature is the high velocity (--110 km s$^{-1}$) ``extension'' of the
southwest lobe of the CND seen by \citet{gus87} and \citet{wri00}.
The northern two-thirds of the streamer may show some faint HCN(1-0)
emission, but they do not form a single feature.

The lack of HCN(1-0) emission from the brightest NH\3(3,3) features is
due in part to self-absorption of HCN(1-0) in the GMCs as well as
absorption of HCN(1-0) along the line of sight \citep{wri00}.
Position-velocity cuts by \citet{wri00} along the southern streamer
and 50 km s$^{-1}$ cloud (see Figure 13 in \citet{wri00}) show strong
self-absorption of HCN(1-0) in these features.  Self-absorption of
HCN(1-0) could weaken the intensity of the GMCs and result in
apparently enhanced emission from the CND, where wide line widths make
HCN(1-0) less susceptible to self-absorption.  The NH\3 is less
affected by absorption because the transitions are of higher energies
and are more optically thin than the HCN(1-0) lines.

\subsection{{\it Comparison to HCN(3-2)}}

Observations of the higher HCN ($J=$3-2) transition by \citet{mar95}
show good agreement with NH\3(3,3) and support the idea that HCN(1-0)
is highly affected by self-absorption and absorption along the
line-of-sight.  Channel maps by \citet{mar95} of HCN(3-2) emission
show strong emission at the location of both the 20 and 50 km s$^{-1}$
clouds in addition to the CND (see Figure 2, \citet{mar95}).  The same
figure also shows features to the west of the CND that agree
kinematically and spatially with NH\3 emission from the western
streamer.  Emission to the southwest begins at --110 km s$^{-1}$ and
continues to the north where it reaches positive velocities of +50 km
s$^{-1}$.  The extension of the 20 km s$^{-1}$ cloud towards the
southwestern lobe of the CND, seen at $17\h45\m36\s.5$,
$-20\dg02'00''$ in Figure \ref{33.fig}, is observed at 10 km s$^{-1}$.
Emission from the southern streamer is seen in the 30 km s$^{-1}$
channel and is just as strong as emission from the CND.  The 50 km
s$^{-1}$ and 20 km s$^{-1}$ GMCs also dominate the channel maps that
correspond to their central velocities.  The northern ridge is easily
observable in HCN(3-2) as the ridge to the north of the CND in the
--10 km s$^{-1}$ channel of \citet{mar95}.  In conclusion, the
NH\3(3,3) and HCN(3-2) images agree on every main feature including
the southern streamer, northern ridge, and western streamer which are
completely absent in the HCN(1-0) image.  This indicates that images
made with HCN(1-0) must be carefully studied to determine the effects
of absorption and excitation.

\subsection{{\it NH\3 as a Column Density Tracer}}

The colorscale in Figure \ref{pbcor} shows a 1.2 mm continuum image
taken by \citet{zyl98} using the IRAM 30 m telescope.  This 1.2 mm
image traces thermal dust emission in the region with additional
free-free emission in the mini-spiral.  Overlaid on the continuum
emission is a primary beam corrected NH\3 (3,3) velocity integrated
map.  The primary beam correction was performed by dividing the
velocity integrated (3,3) emission by the gain of the interferometer
at each point in the map.  Only points with a gain $\ge$15\% are
plotted in Figure \ref{pbcor}.  Primary beam correction results in a
map with varying noise characteristics across the field.  Although the
rms noise of this image near the field center is still $\sim$0.33 Jy
beam$^{-1}$ km s$^{-1}$, the noise is scaled up by the reciprocal of
the gain and is $\sim$2.2 Jy beam$^{-1}$ km s$^{-1}$ at the map edge.

There is a striking correspondence between the thermal dust and NH\3
(3,3) emission.  The southern streamer as well as SE1 and SE2 are
clearly visible in both NH\3 (3,3) and 1.2 mm dust.  The 20 km
s$^{-1}$ GMC is the bright dust feature at $17\h45\m41\s$,
$-29\dg02'30''$ and is well traced by NH\3 (3,3) to the edge of the
mosaic. The 50 km s$^{-1}$ cloud can be seen as the bright dust
emission to the northeast from $17\h45\m51\s$, $-29\dg00'15''$ to
$17\h45\m51\s$, $-28\dg59'00''$.  This GMC is also close to the edge
of the NH\3 image and can be seen as the bright NH\3(3,3) feature on
the northeastern edge of the mosaic.  There is also striking agreement
along the ``northern ridge.''  This feature is clearly present in the
dust image from $17\h45\m44\s$, $-28\dg59'00''$ to $17\h45\m47\s$,
$-28\dg58'00''$ where it intersects the 50 km s$^{-1}$ cloud.  Both
tracers show little or no emission in the northwest quadrant of the
image.  There is also a lack of NH\3(3,3) and thermal dust emission in
the cavity of Sgr A East covering $17\h45\m48\s.5$ to $17\h45\m44\s$
and $-29\dg01'00''$ to $-28\dg58'45''$.  The degree of correlation
between NH\3 and dust emission in the CND is difficult to determine
due to the free-free emission that dominates the dust image near Sgr A
West.


Comparison of the 1.2 mm dust map to a 350 $\mu$m map by \citet{dow99}
shows that the dust emission is approximately constant over this
wavelength range. The similarity of the two maps implies that the 1.2
mm dust traces column density and is not strongly correlated with
temperature in the region.  The dust image has a beam size of
$\sim$11$''$, and as a single dish map it detects highly extended
emission.  Considering the lack of zero-spacing information in the
NH\3 data (shortest baseline 35 m), the agreement between the two maps
is compelling, suggesting that NH\3 is a relatively unbiased tracer of
column density in the Galactic Center region.

A few features seen in the NH\3(3,3) image are not observed in the
thermal dust image.  The ``western streamer'' is not strongly
correlated with the dust, and there is almost no dust emission in the
upper two-thirds of this streamer.  Line ratios of more than 1 in this
region (see Figure \ref{temp.fig}) indicate that the western streamer
is heated to $T_R\simgt30$ K.  The narrow stream of gas connecting the
northern ridge to the CND is also not strongly correlated with the
dust and shows similar hints of high line ratios in Figure
\ref{temp.fig}.  The heating and absence of dust in these features
indicate that they may be located physically closer to Sgr A* or may 
originate in a different way than the streamers that contain dust.

\subsection{{\it Interactions with Sgr A East}} 

The velocity integrated NH\3 (3,3) emission is overlaid on 6 cm
continuum emission \citep{yus87} in Figure \ref{sgeast.fig}.  The arms
of the Sgr A West mini-spiral can be seen in blue in the center of the
image.  The eastern edge of Sgr A East extends to $\sim17\h45\m50\s$
and the western edge is spatially coincident with the western edge of
the CND and mini-spiral. Sgr A East is expanding with an energy more
than an order of magnitude greater than a typical supernova remnant
\citep{mez89,gen90}.  The NH\3 (3,3) emission closely follows the
outer edge of Sgr A East.  The shell is impacting the 50 km s$^{-1}$
cloud in the east \citep{gen90,ho91,ser92, zyl99} where material forms
a ``ridge'' on the western edge of the cloud.  This ridge is seen in
NH\3(3,3) from $17\h45\m51\s$, $-29\dg00'15''$ to $17\h45\m47\s$,
$-28\dg59'00''$.  The northern ridge and western streamer also lie
along the outside edge of Sgr A East.  Therefore, it appears that Sgr
A East is expanding into material on all sides of the CND.  We observe
multiple connections between the CND and the ridges on the outside
edge of Sgr A East (see Section 5).  These connections are a strong
indication that the CND and Sgr A East are in close proximity to each
other.

The southern edge of Sgr A East is approximately coincident with the
two southern clouds and the southern streamer.  Since the ``20 km
s$^{-1}$'' giant molecular cloud (which appears to be the source of at
least the southern streamer \citep{ho91,coi99,coi00}) is thought to be
located in front of Sgr A East \citep{coi99,coi00}, Sgr A East may be
expanding into the back of these filaments.  In addition, there is
evidence that a SNR centered at $\Delta_{\alpha}\sim 80'',
\Delta_{\delta}\sim -120''$ \citep[and references therein]{coi00} may
be impacting Sgr A East along its southeastern edge.  This could
account for the 1720 MHz OH masers detected by \citet{yus99} on the
boundary between this SNR and Sgr A East.  The interaction of these two expanding shells could also produce filamentary structures like SE1, SE2, and the southern streamer.

Figure \ref{sgeast.fig} is also useful for studying the gap in NH\3
(3,3) emission in the northern part of the CND at the location of the
northern arm of the mini-spiral.  As discussed in Section 4.1, this
gap is also seen in HCN (1-0) emission.  The lack of NH\3 (3,3)
emission in addition to the lack of HCN(1-0) emission and the
observation of 1720 OH masers in this gap supports the idea that the
northern arm of the mini-spiral originates outside the CND and is
crossing over or through the CND on its way to the Galactic Center
\citep{wri00}.

\section{Connections between the CND and GMCs}

Connections between the CND and features at larger distances such as
the 20 and 50 km s$^{-1}$ cloud can provide an explanation for the
origin of the clouds that compose the CND.  A connection must show more
than morphological evidence to be considered a physical gas flow.  The
presence of a velocity gradient along a streamer indicates that the
gas is being gravitationally affected by the central gravitational
field.  Line ratios can also determine whether gas is heated as it
approaches the nucleus, thus providing a way to discriminate
against chance projections of clouds along the line-of-sight.
Finally, as the gas intersects clouds in the CND, the line widths
should broaden.  Streamers with these qualities are
likely candidates for flows along which gas is transported towards the
Galactic Center.

We detect at least three physical connections between the CND and gas
at larger distances.  These connections originate in the southern
streamer, SE1, the northern ridge, and the western streamer and may
represent inflow of material towards the nucleus.  Figure \ref{pvpos}
shows the position of six position-velocity cuts along features which
appear to be connecting the material to the CND overlaid on NH\3(3,3)
emission.  The position velocity diagrams for NH\3 (1,1), (2,2) and
(3,3) are shown in Figures \ref{11pv}, \ref{22pv}, and \ref{pv}.  In
each cut, position ``0'' corresponds to the labeled end of the cut.

In the velocity integrated images, the brightest emission is at 30-50
km s$^{-1}$ mainly from the large GMCs.  When performing MEM
deconvolution, channels with bright emission may have ``blooming''
where plateaus of low-level flux are added around strong features.
Artifacts of blooming were seen in NH\3(3,3) position-velocity
diagrams as narrow lines centered at $\sim40$ km s$^{-1}$ seen along
most of the cut. To test the effects of blooming, we created CLEANed
data cubes.  The bright emission is identical in both diagrams, but
the narrow ridges centered at 45 km s$^{-1}$ disappear in the CLEANed
diagrams.  To be conservative, we present the CLEANed diagrams in
Figures \ref{11pv}, \ref{22pv}, and \ref{pv}.

\subsection{{\it The Southern Streamer}}

The southern streamer has previously been observed to connect the 20
km s$^{-1}$ cloud to the southern lobe of the CND \citep{coi99,coi00}.
Position velocity cut $a$ in Figures \ref{11pv}, \ref{22pv}, and
\ref{pv} show emission along the southern streamer.  A small velocity
gradient from 30 km s$^{-1}$ at a position of 40$''$ to 40 km s$^{-1}$
at 70$''$ can be seen as gas approaches the CND.  With higher velocity
resolution, \citet{coi99} were able to detect a comparable shift of
$\sim$ 10 km s$^{-1}$ in this region of the southern streamer (see
Fig. 8, cut $a$ in \citet{coi99}).  In addition, the line is seen to
broaden dramatically when the gas reaches the CND.  At a position of
90-100$''$, NH\3(3,3) emission is seen over a velocity range of almost
100 km s$^{-1}$.  This broadening is also seen in NH\3(2,2) and to a
lesser degree in NH\3(1,1).  The peak of the emission is located at
$\sim30$ km s$^{-1}$ while broadening is seen in the lowest contours.
The broadening in all three transitions seems to be biased towards
lower velocities and ranges from --70 to +50 km s$^{-1}$ instead of
being symmetric on both sides of the line center.

An increase in the NH\3(2,2) to (1,1) line ratio as the streamer
approaches the CND can be seen in Figure \ref{temp.fig} and indicates
that gas is being heated as it approaches the nucleus.  This heating
could result from shocks as the gas merges with the CND or from
high-energy photons escaping from the nuclear region.  The NH\3(3,3)
hyperfine satellite components may be seen bracketing the strong
central component.  Asymmetric broadening of the main component may
explain why the satellite lines do not appear to have equal intensity
on both sides of the main component in position-velocity cut $a$.

In Figure \ref{pv}$a$, NH\3(3,3) emission at 30 km s$^{-1}$ and
110$''$ to 140$''$ appears to come from inside the CND.  NH\3(3,3)
emission at 160$''$ is associated with the northwestern side of the
CND.  In NH\3(1,1) and (2,2) emission, however, the southern streamer
stops sharply at the CND.  This abrupt end is used by \citet{coi99,
coi00} as evidence that the southern streamer is interacting with the
CND.  The small amount NH\3(3,3) gas inside the CND is probably hot
since there is no emission in NH\3(1,1) or (2,2).

\subsection{{\it The Northern Ridge}}

Faint NH\3 emission can be seen in Figures \ref{11.fig} and
\ref{33.fig} connecting the northern ridge to the northeastern lobe of
the CND.  Position-velocity cut $b$ in Figure \ref{pv} shows the
kinematic structure of this feature in NH\3(3,3).  Gas in the northern
ridge at a position of 30$''$ has a velocity of --10 km s$^{-1}$ and
connects to the CND with a smooth velocity gradient of 0.6 km
s$^{-1}$arcsec$^{-1}$.  This large velocity gradient covers 110$''$ (4
pc) and reaches the CND at 60 km s$^{-1}$.  Position velocity cuts in
NH\3(1,1) and (2,2) show the same kinematic signature for this
streamer connecting the northern ridge to the CND.  The positive velocity
increase as the gas approaches the CND places the northern ridge in
front of the northeastern part of the CND.  In addition, the low
velocity of this streamer ($\simlt 60$ km s$^{-1}$) makes it likely to
be gravitationally bound by the nucleus.  Since the velocity increases
as projected distance decreases, we can eliminate a circular orbit for
this cloud.

The northern ridge has high NH\3(2,2) to (1,1) line ratios and may
have been impacted and disrupted by Sgr A East (see section 4.3).
Line ratios for most of the northern ridge and connecting stream are
$\simgt 1$.  Position-velocity cut $b$ can be used to better estimate
the line ratios. At position 40$''$, the ratio of NH\3(2,2) to
NH\3(1,1) emission is $\sim0.7$.  This ratio increases to $\sim1$
where the streamer intersects the CND (position 140$''$).  Assuming
that the streamer is optically thin along its entire length, this
increase in NH\3(2,2) to (1,1) line ratios corresponds to a
temperature increase from $\sim25$ K to $\sim30$ K.

Although the upper part of the northern ridge is well traced in dust,
there is little dust emission along the connection to the CND (see
Figure \ref{pbcor}).  The northern ridge appears to have been swept
up by the expansion of Sgr A East.  Some of the disrupted gas may be
falling towards the CND and it is possible that dust in the gas has
been destroyed by photons from the nucleus.

\subsection{{\it SE1}}

Although it is faint, SE1 and the CND are connected in all three
velocity integrated images.  Analysis of the kinematic data shows a
strong connection between SE1 and northeastern lobe of the CND.  This
connection can be easily seen in Figure \ref{pv}$c$.  To the south,
SE1 has a velocity of 40 km s$^{-1}$ at 50$''$.  The velocity
increases to the north until it reaches 50 km s$^{-1}$ at 90$''$.  A
connection to the CND at 50 km s$^{-1}$ extends from 110$''$ to
140$''$ where it intersects the CND at the same velocity.  SE1 is also
connected to the CND by emission at 50 km s$^{-1}$ in NH\3(1,1) and
(2,2).  The connection is narrow and weaker than emission from either
SE1 or the CND.  It may be weak due to the intrinsic clumpiness of the
clouds, or it also may have been disrupted by an interaction with
another cloud.  

Kinematic evidence from HCN (1-0) indicates that the northern half of
the CND may be on a different orbit than the rest of the ring
\citep{wri00}.  The independence of this feature from the rest of the
CND would also account for the different inclination angles observed
for the two halves of the CND.  Given the strong connection between
the CND and SE1, it appears that some of the gas in the northern lobe
originated in SE1 and is not part of a coherent rotating ring.

\subsection{{\it The Western Streamer}}

Position velocity cut $d$ in Figure \ref{pv} shows a striking velocity
gradient of 1 km s$^{-1}$arcsec$^{-1}$ along the entire 150$''$ (6pc)
length of the western streamer.  The southern end of the streamer at
60$''$ has a velocity of --70 km s$^{-1}$ while the northern end has
high velocity emission at +90 km s$^{-1}$ at 220$''$.  The same
velocity gradient is observed in NH\3(1,1) and (2,2) in
position-velocity diagrams \ref{11pv}$d$ and \ref{22pv}$d$.  The large
velocity gradient along the length of the cloud could be due to
intrinsic rotation or the cloud could be orbiting the nucleus.  If we
assume a circular orbit for the western streamer at a distance of 2$'$
(4.6 pc) the observed velocity gradient is consistent with a keplerian
orbit around a central mass of 10$^7$ M$_\odot$ (including Sgr A* and
the stellar population, see \citet{hal96}) inclined by $\sim30^\circ$
to the line of sight.  Additionally, the impact of Sgr A East could
enhance the gradient along this streamer.

The western streamer is not seen in 1.2 mm dust emission (see Figure
\ref{pbcor}).  The dust may have been destroyed by photons from the
nucleus or by interactions with Sgr A East.  As seen in Figure
\ref{sgeast.fig}, the curve of the western streamer follows the edge of
Sgr A East.  It is possible that this gas originated closer to the
nucleus and was pushed outward.  In this scenario, any dust that was
in the gas was removed when it was close to the nucleus.

Three narrow filaments appear to connect the western streamer to the
CND in the velocity integrated maps.  The southern-most projected
connection, at $17\h45\m36\s.5$, $-29\dg02'00''$, is visible in
position-velocity cut $d$ at 20$''$ with a velocity of 0 km s$^{-1}$.
This cloud is the same extension of the 20 km s$^{-1}$ cloud towards
the southwestern lobe of the CND that was seen by \citet{coi99,coi00}
in NH\3(1,1) and (2,2).  Kinematically, it is associated with the 20
km s$^{-1}$ cloud and not the western streamer.  Although the
extension is spatially connected to the CND in NH\3(3,3), HCN(1-0) and
dust emission (see Figures \ref{hcn.fig} and \ref{pbcor}), there is
no kinematic evidence for a physical connection between the 20 km
s$^{-1}$ cloud and this south-western part of the CND which has
typical velocities of --110 km s$^{-1}$.  The use of  CLEANed data
for the position-velocity diagrams makes it difficult to see faint,
extended connections between features.  It is possible that features
which show morphological connections but no obvious kinematic
associations with the CND are connected to the CND by faint and
extended gas.

The second possible connection between the western streamer and the
southwest lobe of the CND is at $17\h45\m36\s.5$, $-29\dg00'55''$.  In
position velocity cut \ref{pv}$e$ the bright clump at 20$''$, centered
at -20 km s$^{-1}$, is the western streamer.  Emission from the CND is
is seen from 80$''$ to 180$''$.  In the western-most part of the CND,
there is bright emission at the same velocity as the western streamer
(80$''$), but any connection is tenuous.  The line widths are
extremely high in this region (FWHM up to 50 km s$^{-1}$) indicating
turbulence.  NH\3(1,1) and (2,2) show the same kinematics.

In NH\3(3,3), the southwestern lobe of the CND at 80-100$''$ shows two
bright clumps centered at $\sim80$ km s$^{-1}$ and $\sim-10$ km
s$^{-1}$.  In NH\3(1,1) there is prominent emission at $\sim-100$ km
s$^{-1}$ at 110$''$ which is either part of the ``negative velocity
lobe'' of the CND observed in HCN(1-0) or high velocity gas at the
nucleus.  The features at --10 and +80 km s$^{-1}$ are more prominent
in NH\3 and do not fit a rotation pattern.  In particular, the feature
at --10 km s$^{-1}$ has a high line width which extends
over 60 km s$^{-1}$, especially in NH\3(2,2).

Position-velocity cut $f$ follows northern-most possible connection to
the CND at $17\h45\m37\s$, $-29\dg00'00''$.  The western streamer is
at a velocity of 30 km s$^{-1}$ at position of 40$''$.  There is no
obvious connection to the CND which is at 70 km s$^{-1}$ and has a
velocity gradient going to 30 km s$^{-1}$ towards the east.  

The two northern projected connections between the CND and the western
streamer are seen as filaments in 6 cm continuum emission.  The
filamentary structures in the continuum emission are intriguing and
may be the result of supernova remnants or expansion of Sgr A
East through the CND.

Overall, we detect three physical connections to the CND.  We confirm
the presence of the southern streamer which connects the 20 km
s$^{-1}$ cloud to the CND.  The northern ridge is also connected to
the CND with a velocity gradient of 0.6 km s$^{-1}$ arcsec$^{-1}$
spanning 110$''$ (4 pc).  SE1 extends northwards and kinematically
connects to the eastern lobe of the CND.  The western streamer shows a
velocity gradient of 1 km s$^{-1}$ arcsec$^{-1}$ along a length of
150$''$ (6pc), but we do not see any definite physical connections to
the CND.

\section{A High Velocity Cloud Near the Nucleus}
There is significant absorption in NH\3(1,1) and (2,2) where cut $e$
passes close to Sgr A* at 120$''$.  The NH\3(3,3) diagram, however,
has little absorption and instead shows emission with a coherent
velocity gradient from +80 km s$^{-1}$ at 95$''$ to 0 km s$^{-1}$ at
145$''$.  This feature is within 2 pc in projected distance from Sgr
A*.  The positive velocity lobe of the CND is seen at 170$''$ and is
not a continuation of the gradient.  A similar gradient is seen as cut
$f$ passes through the interior of the CND (70-130$''$).  Unlike cut
$e$, the velocity gradient of gas near Sgr A* in cut $f$ is not
constant.  The velocity decreases from 75 km s$^{-1}$ at 75$''$ to 50
km s$^{-1}$ at 90$''$ and then is approximately constant at 35 km
s$^{-1}$ from 100-115$''$.  At 125$''$, there is emission from --50 km
s$^{-1}$ to +40 km s$^{-1}$.  Although not shown in this paper, the
same feature is seen over a range of position angles from 60-100$\dg$.
The feature must contribute to NH\3(3,3) emission seen close to Sgr A*
in the velocity integrated image (see Figure \ref{33.fig}), but the
resolution of the image is too coarse to outline its morphology.  The
observed velocity gradient may be the result of a cloud being tidally
stripped by the central gravitational potential or may be a feature
orbiting the black hole in the opposite sense as the CND.  At the
position of Sgr A* (125$''$ in $e$ and 110$''$ in $f$), the velocity
is centered at 30 km s$^{-1}$ making it unlikely that this feature is
a counter-rotating inner disk around Sgr A*.  If we assume this
feature is within 2 pc of the Galactic Center, then its maximum
velocity (100 km s$^{-1}$) is well below the escape velocity for a
gravitational potential of $10^7$M$_\odot$ and the cloud is likely to
be gravitationally bound to the nucleus.  However, considering
projection effects and without the spatial morphology of this cloud,
it is not possible to uniquely determine an orbit for this feature.

\section{Conclusion}
VLA mosaics of the central 10 pc of the Galaxy in NH\3(1,1), (2,2),
and (3,3) allow a detailed study of gas on all sides of the CND.  The
velocity integrated images show many filamentary features including
the southern streamer, northern ridge, and western streamer as well as
a high velocity cloud within 2 pc of the Galactic Center.  There is
a high correlation between NH\3(3,3) emission and emission from dust
and HCN(3-2), suggesting that NH\3(3,3) traces column density.  Line
ratios of NH\3(2,2) to (1,1) emission indicate that the CND and
western streamer are hotter than GMCs which lie at a projected
distance of $\sim10$ pc.  The CND is well-traced in NH\3(3,3)
although the relative intensity of features differs from HCN(1-0).
Outside the CND, the 50 km s$^{-1}$ cloud, northern ridge and western
streamer are located around the edge of Sgr A East and appear to have been
swept up by the expanding shell.  Many of the filamentary features
including the northern ridge, southern streamer and SE1 are
kinematically connected to the CND indicating that GMCs are feeding
the nucleus along many paths.

\acknowledgements{The National Radio Astronomy Observatory is a
facility of the National Science Foundation operated under cooperative
agreement by Associated Universities, Inc.  ALC is supported by a NSF
Graduate Research Fellowship.  We would like to thank M. Wright for
the use of the HCN(1-0) image, R. Zylka for the use of the 1.2 mm dust
image and D. Shepherd for helpful discussions.  The 6 cm continuum
image was obtained from the ADIL library.}

\newpage
\title{CAPTIONS}

\noindent
For full resolution figures, see \url{http://cfa-www.harvard.edu/$\sim$rmcgary/SGRA/nh3\_figures.html}

\figcaption[McGary.fig1.eps]{Velocity integrated NH\3 (1,1) emission.  The rms noise of the map is $\sigma_{11}=0.28$ Jy beam$^{-1}$ km s$^{-1}$ and the contour levels are in intervals of $3\sigma_{11}$.  In all three velocity integrated maps a primary beam response of 10, 30 and 50\% is shown in dashed contours.\label{11.fig}}

\figcaption[McGary.fig2.eps]{Velocity integrated NH\3 (2,2) emission.  The rms noise of the map is $\sigma_{22}=0.30$ Jy beam$^{-1}$ km s$^{-1}$ and the contour levels are in intervals of $3\sigma_{22}$.\label{22.fig}}

\figcaption[McGary.fig3.eps]{Velocity integrated NH\3 (3,3) emission with main features outside the CND labeled.  The rms noise of the map is $\sigma_{33}=0.33$ Jy beam$^{-1}$ km s$^{-1}$ and the contour levels are in intervals of $3\sigma_{33}$.\label{33.fig}}

\figcaption[McGary.fig3b.eps]{Channel maps of NH\3 (3,3) emission with the mean velocity labeled in each map.  Contour levels are 3,6,10,15,23,30,40,55,70 and 90$\sigma$ where $\sigma=0.01$ Jy beam$^{-1}$.  No clipping was performed on these maps and the noise should be equal across the map.\label{chan.fig}}

\figcaption[McGary.fig4.eps]{Ratio of NH\3 (2,2) to (1,1) emission at all points where the (2,2) emission is $>3\sigma_{22}$ and (1,1) emission is $>2\sigma_{11}$.  The 3$\sigma_{22}$ contour of (2,2) emission is also plotted.  A ratio of 0.5 corresponds to $T_R\sim20$K while a ratio greater than 1.5 implies $T_R\simgt50$K (for $\tau\ll0$, \citet{ho83}).\label{temp.fig}}

\figcaption[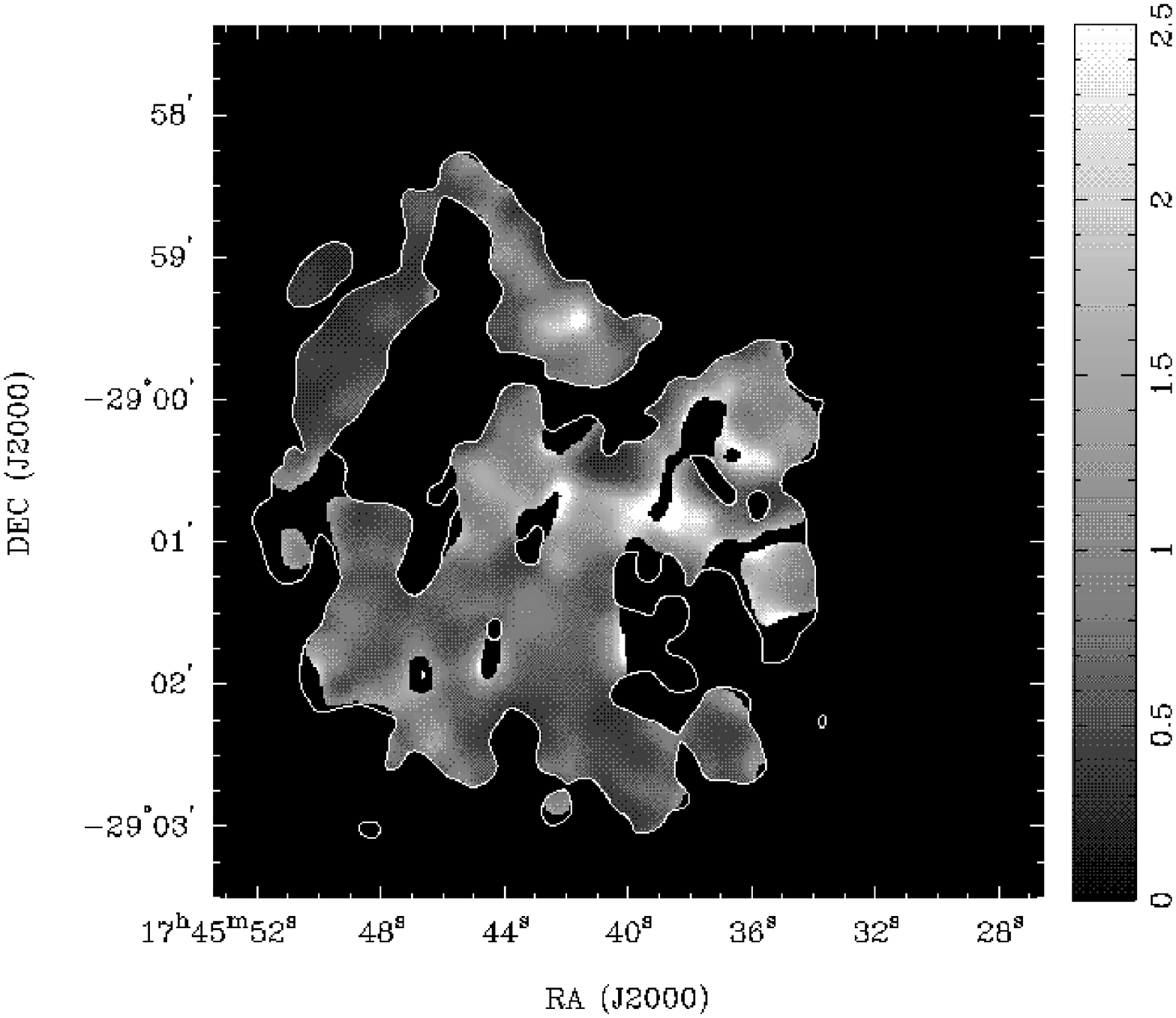]{Comparison of NH\3(3,3) emission (in contours) and HCN(1-0) emission from \citet{wri00} (in color).  The color scale ranges from .05 to 0.5 Jy beam$^{-1}$ km s$^{-1}$. \label{hcn.fig}}

\figcaption[McGary.fig6.eps]{Positions of spectra near the CND.  Spectra A-J are in the same positions as in \citet{wri00}. \label{specpos}}

\figcaption[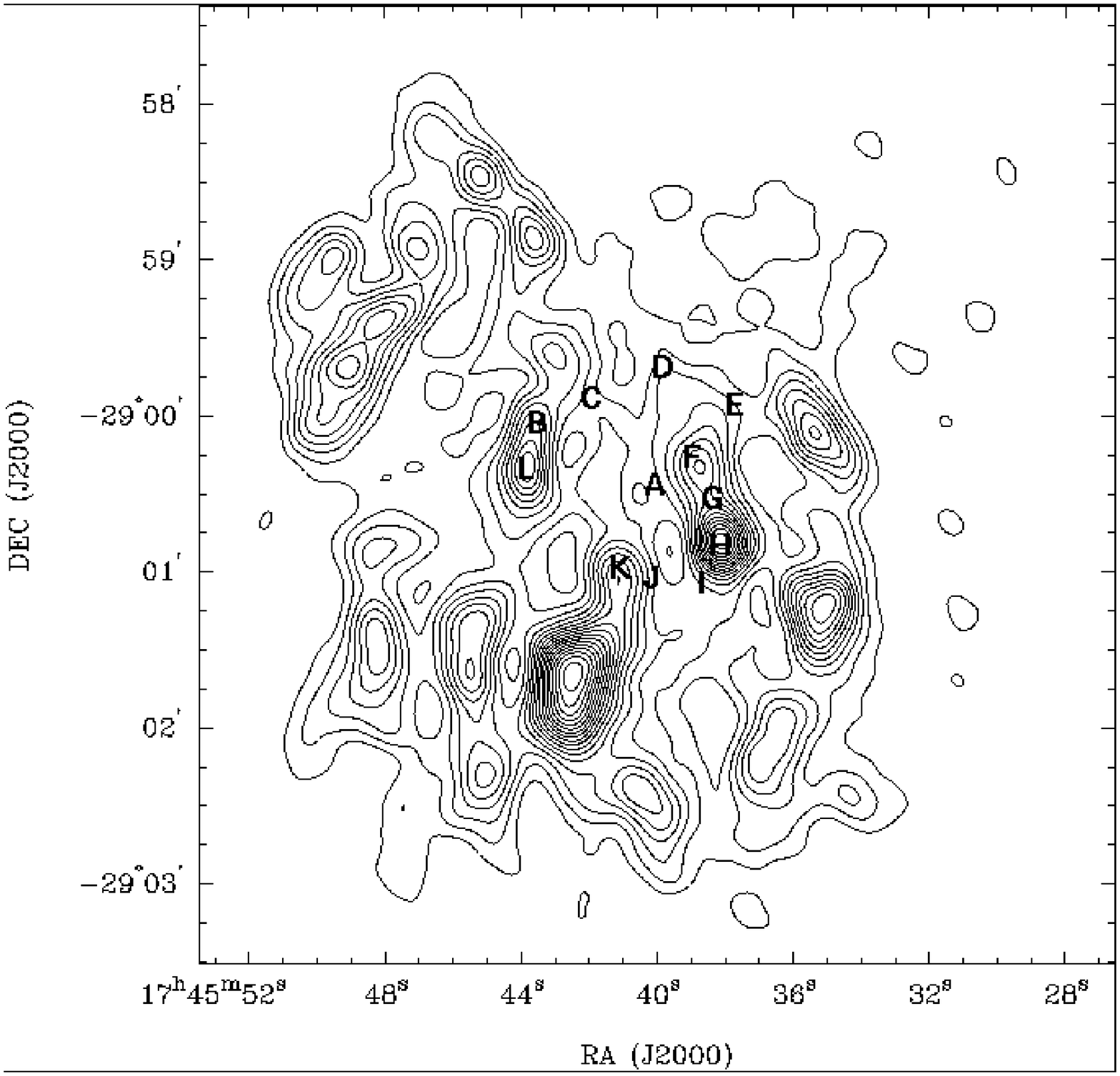]{Spectra from Figure \ref{specpos}.  The NH\3(3,3) spectra are shown in solid lines.  For spectra A-J, the HCN(1-0) spectra from \citet{wri00} are shown as dashed lines.  The vertical scale is in Jy beam$^{-1}$ for the NH\3(3,3) spectra while the HCN(1-0) spectra have been scaled down by a factor of 8 to fit on the plot. \label{spec}}


\figcaption[McGary.fig9.eps]{Primary beam corrected NH\3(3,3) image overlaid on 1.2 mm emission \citep{zyl98}  Contour levels are 3,6,10,15,23,30,40,55,70,90, and 110$\sigma_{33}$ and the color scale runs from .3 to 1 Jy beam$^{-1}$. \label{pbcor}}

\figcaption[McGary.fig10.eps]{Velocity integrated NH\3 (3,3) emission in yellow contours overlaid on 6 cm continuum emission showing Sgr A East \citep{yus87}.  Contour levels are in intervals of $4\sigma_{33}$ and the color scale ranges from 0 to 0.7 Jy beam$^{-1}$.  The positions of the 1720 MHz OH masers from \citet{yus99} are labeled with green error ellipses scaled up by a factor of 15.\label{sgeast.fig}}

\figcaption[McGary.fig11.eps]{Positions of position-velocity cuts on NH\3(3,3) emission. \label{pvpos}}

\figcaption[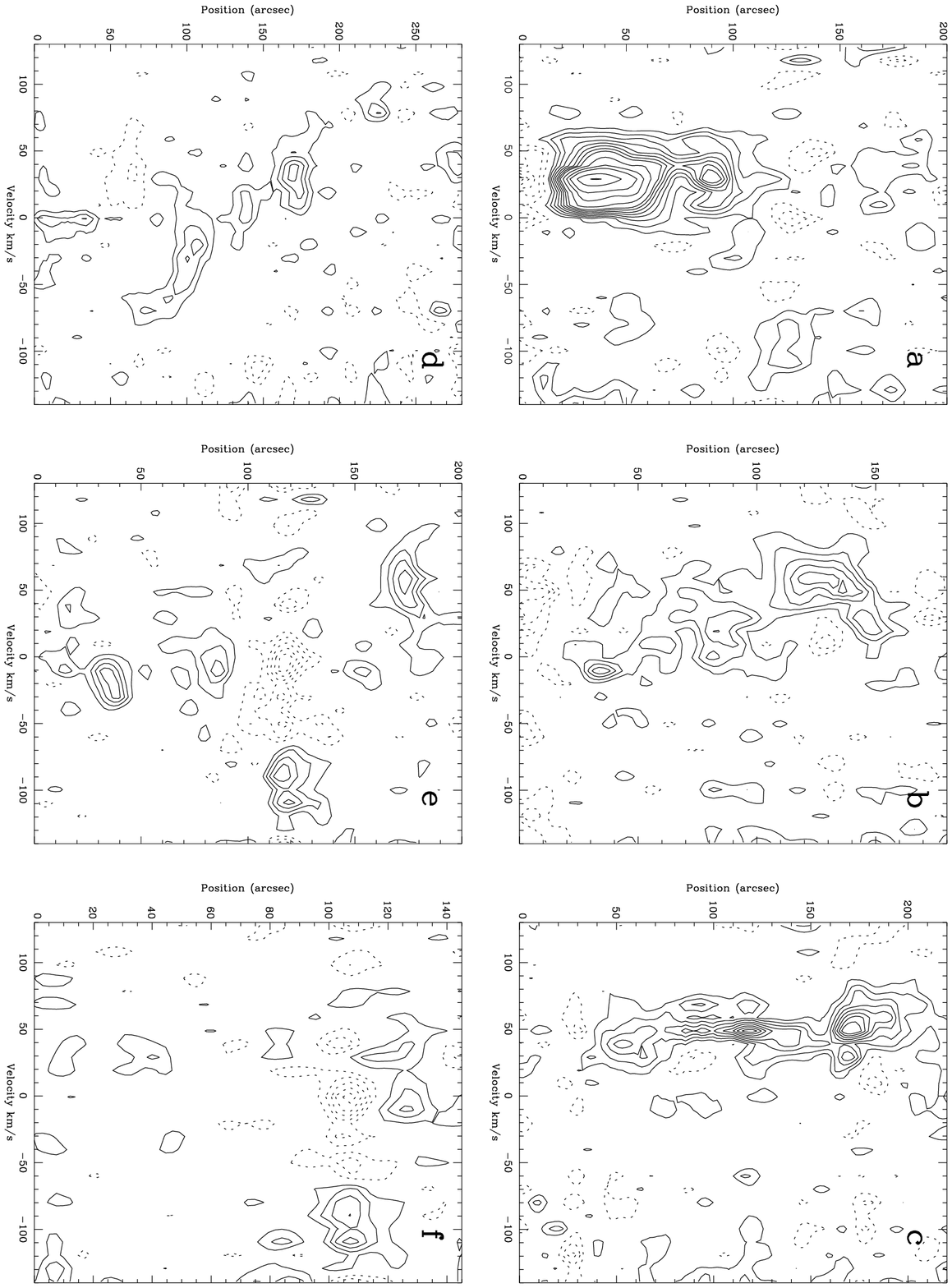]{Position-velocity diagrams for NH\3(1,1) emission.  Position 0 corresponds to the letter label in Figure \ref{pvpos}.  The contour levels are 0.01, 0.02, 0.03, 0.04, 0.05, 0.06, 0.07, 0.08, 0.1, 0.125, 0.15, 0.2, 0.25, 0.3, 0.35, 0.4, 0.45, 0.5, 0.55, and 0.6 Jy beam$^{-1}$.  \label{11pv}}

\figcaption[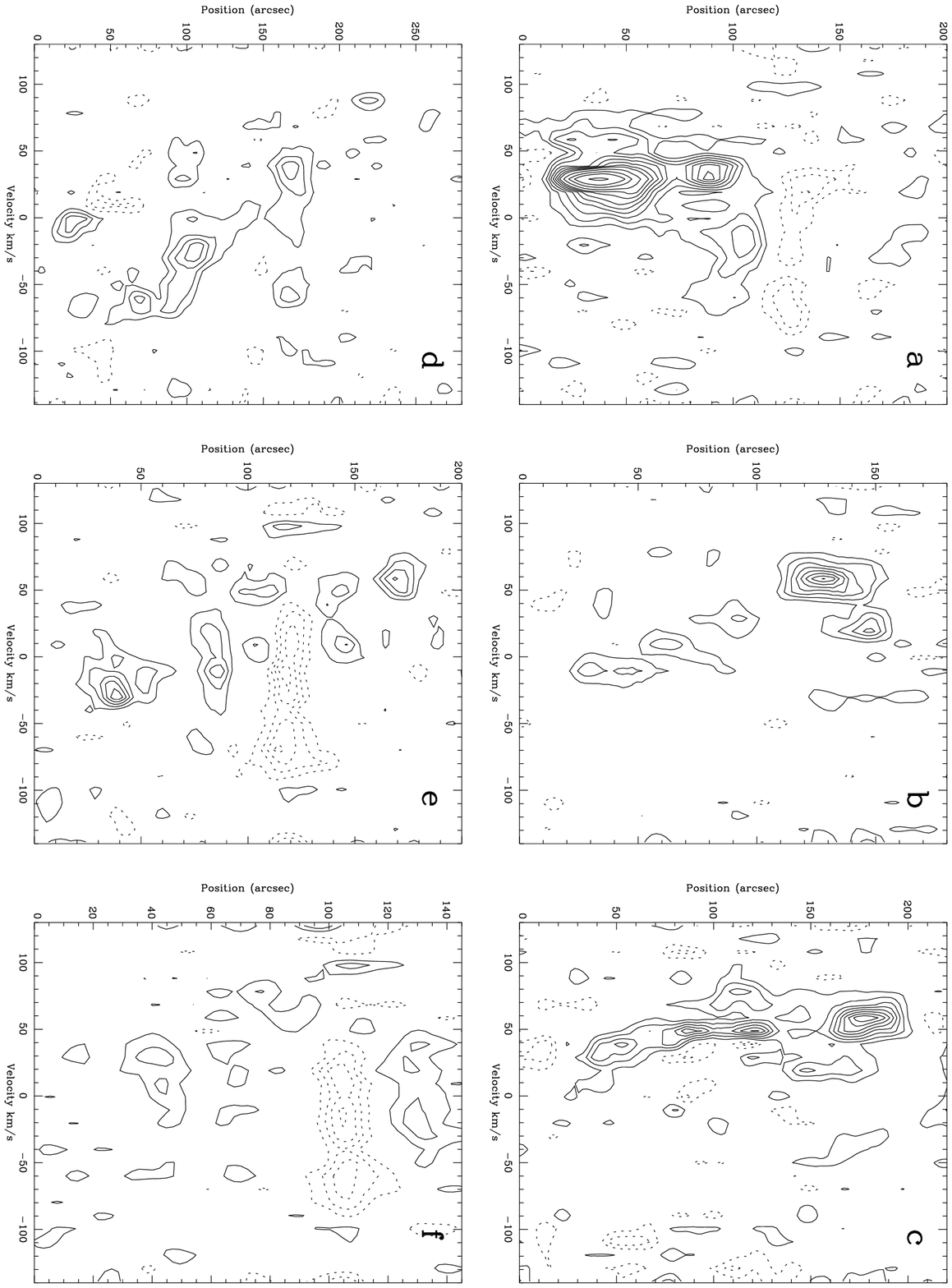]{Position-velocity diagrams for NH\3(2,2) emission.  Position 0 corresponds to the letter label in Figure \ref{pvpos}.  The contour levels are 0.01, 0.02, 0.03, 0.04, 0.05, 0.06, 0.07, 0.08, 0.1, 0.125, 0.15, 0.2, 0.25, 0.3, 0.35, 0.4, 0.45, 0.5, 0.55, and 0.6 Jy beam$^{-1}$.  \label{22pv}}

\figcaption[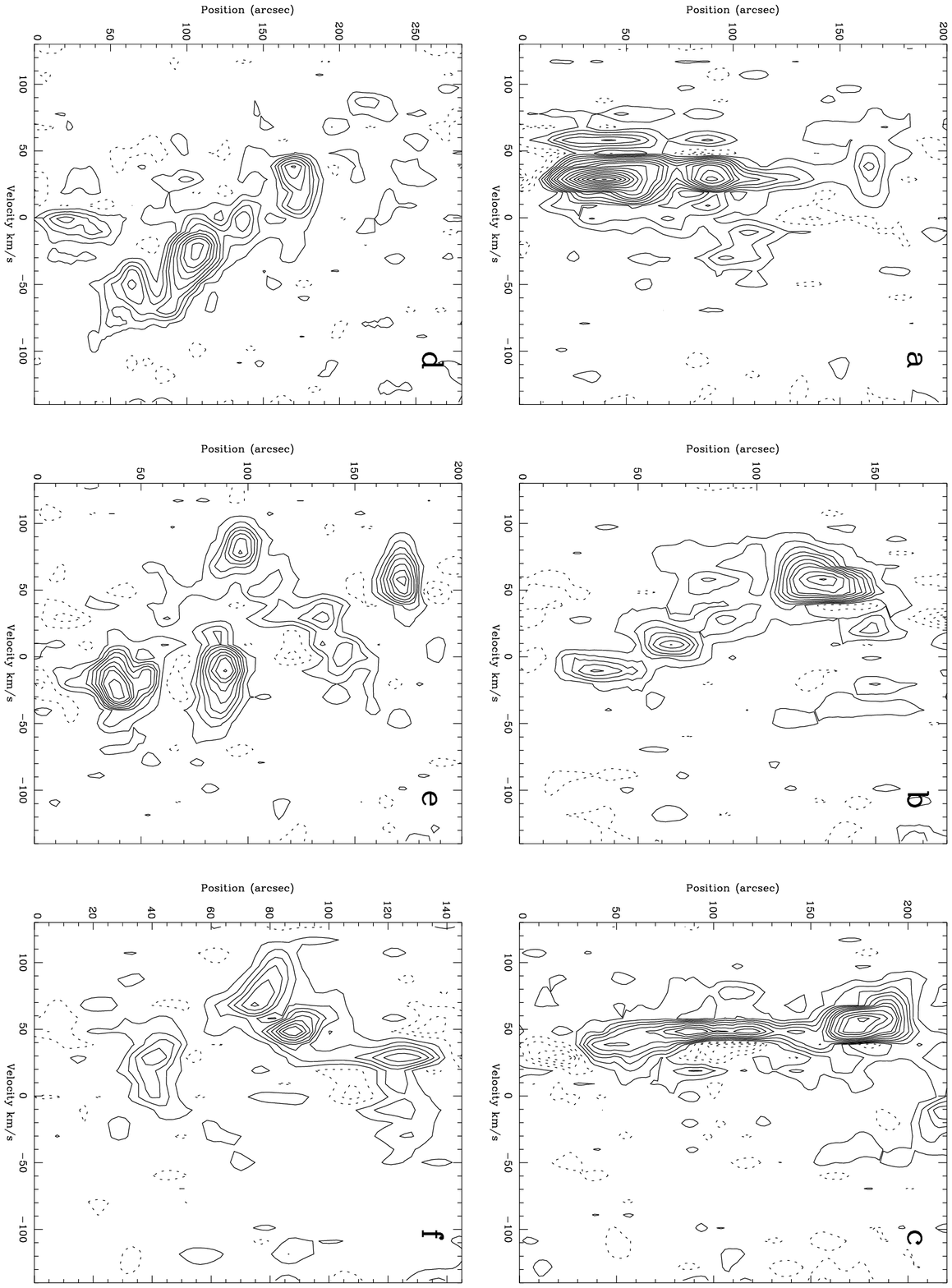]{Position-velocity diagrams for NH\3(3,3) emission.  Position 0 corresponds to the letter label in Figure \ref{pvpos}.  The contour levels are 0.01, 0.025, 0.04, 0.055, 0.075, 0.1, 0.125, 0.15, 0.2, 0.25, 0.3, 0.35, 0.4, 0.45, 0.5, 0.55, and 0.6 Jy beam$^{-1}$.  \label{pv}}

\newpage

\begin{figure}
\figurenum{5}
\plotone{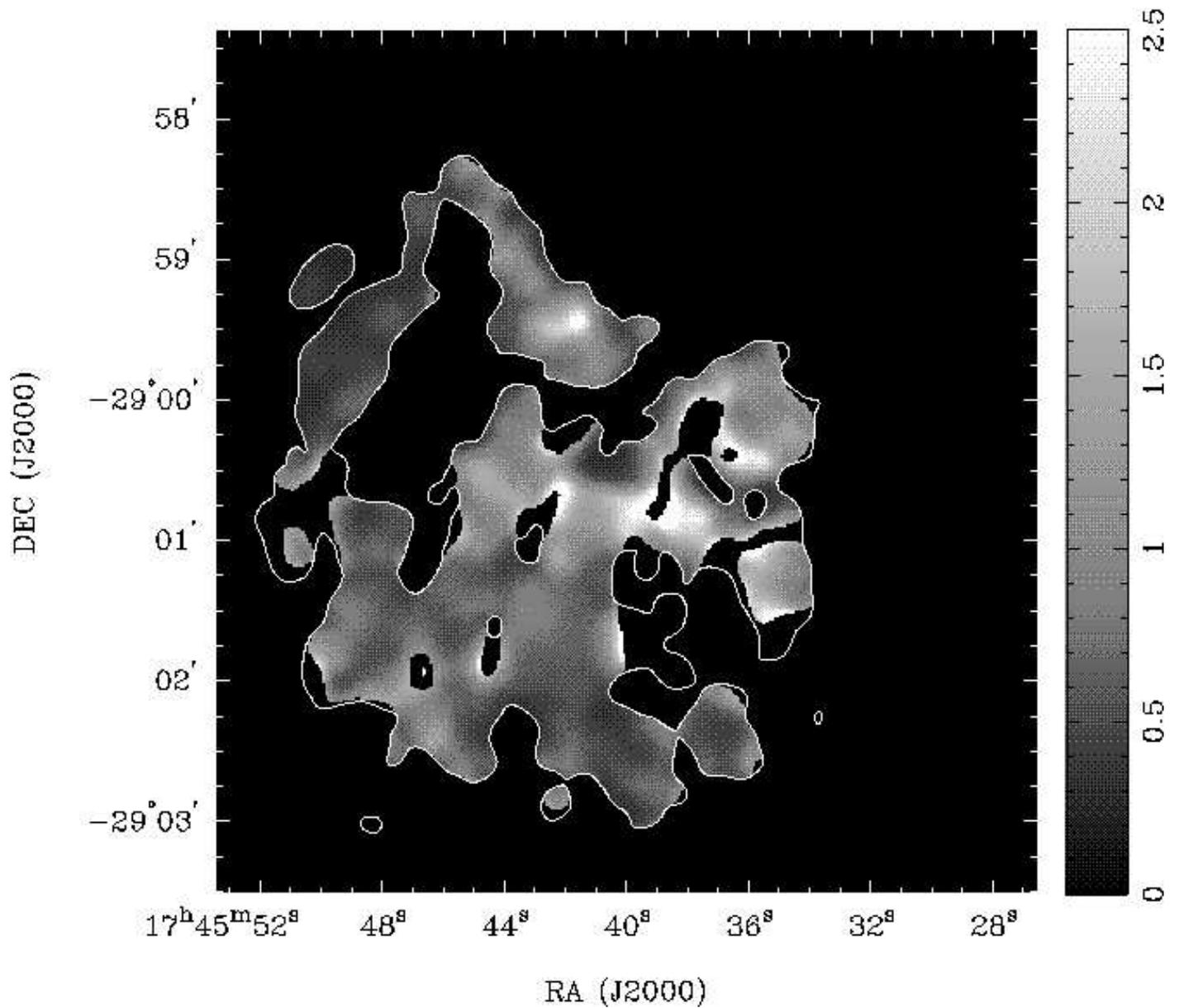}
\caption{Ratio of NH\3 (2,2) to (1,1) emission at all points where the (2,2) emission is $>3\sigma_{22}$ and (1,1) emission is $>2\sigma_{11}$.  The 3$\sigma_{22}$ contour of (2,2) emission is also plotted.  A ratio of 0.5 corresponds to $T_R\sim20$K while a ratio greater than 1.5 implies $T_R\simgt50$K (for $\tau\ll0$, \citet{ho83}).}
\end{figure}

\begin{figure}
\figurenum{7}
\plotone{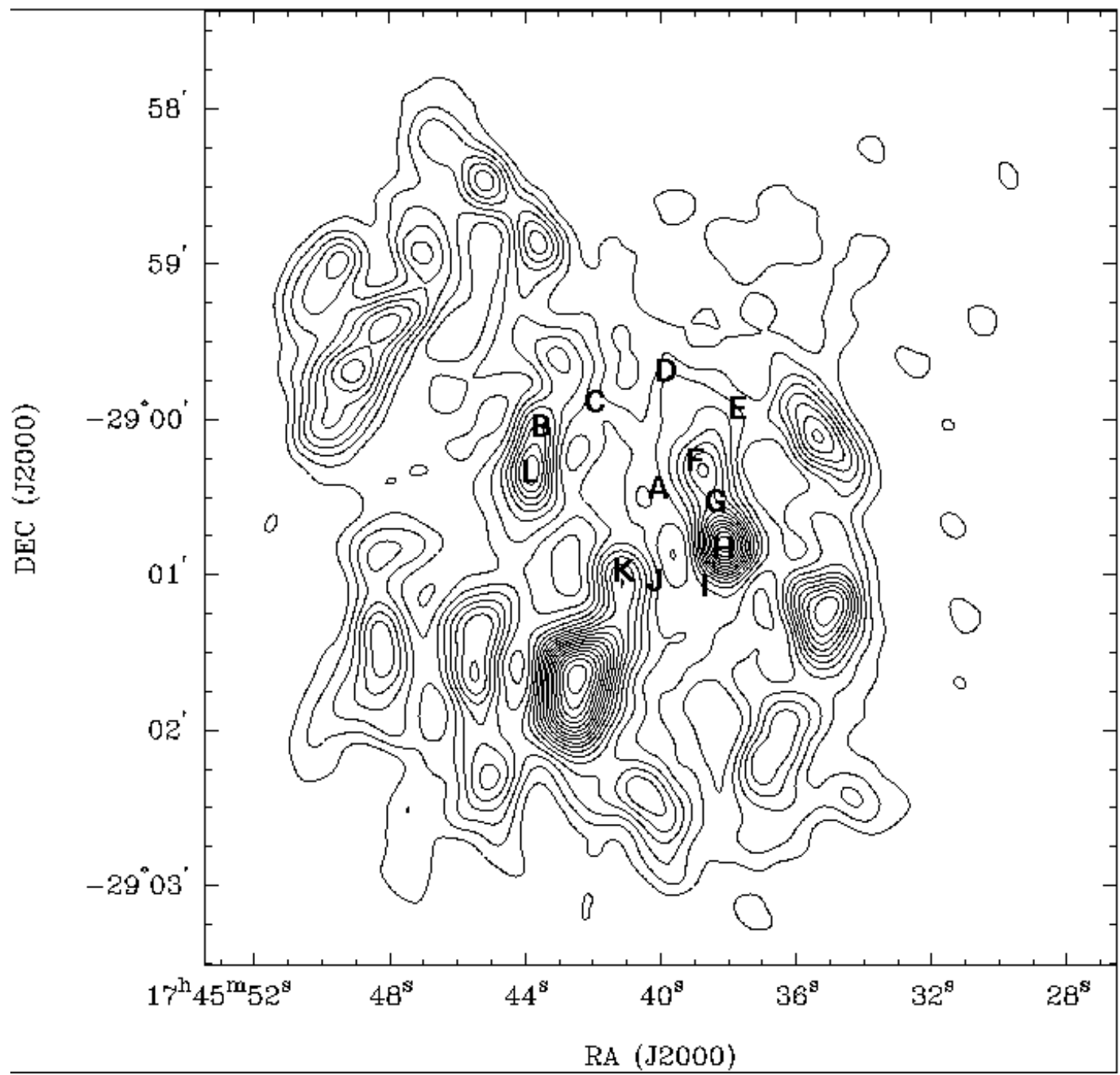}
\caption{Positions of spectra near the CND.  Spectra A-J are in the same positions as in \citet{wri00}. }
\end{figure}

\begin{figure}
\figurenum{8}
\epsscale{0.75}
\plotone{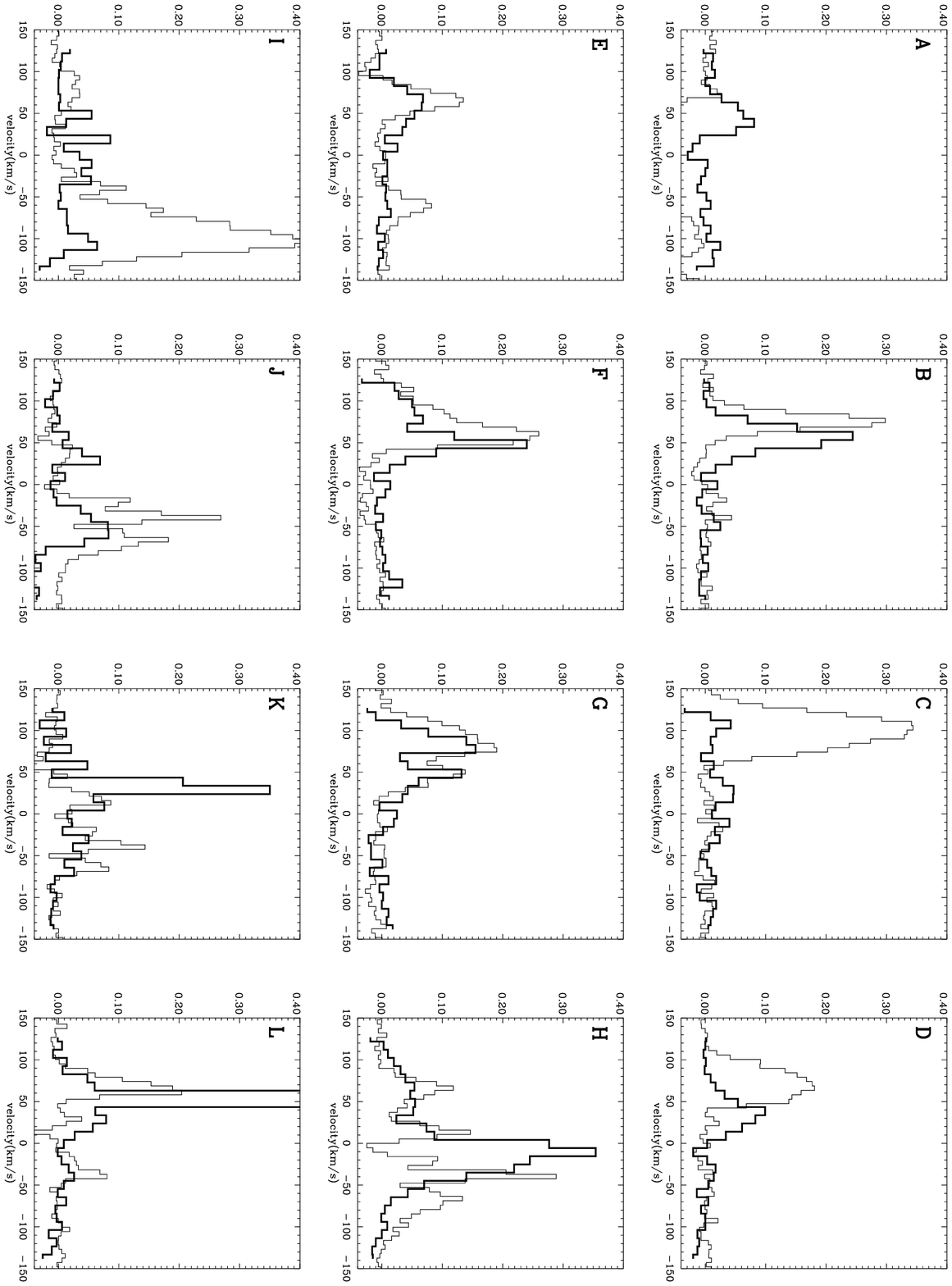}
\caption{Spectra from Figure \ref{specpos}.  The NH\3(3,3) spectra are shown in solid lines.  For spectra A-J, the HCN(1-0) spectra from \citet{wri00} are shown as dashed lines.  The vertical scale is in Jy beam$^{-1}$ for the NH\3(3,3) spectra while the HCN(1-0) spectra have been scaled down by a factor of 8 to fit on the plot.}
\end{figure}

\begin{figure}
\figurenum{12}
\epsscale{0.75}
\plotone{McGary.fig12.eps}
\caption{Position-velocity diagrams for NH\3(1,1) emission.  Position 0 corresponds to the letter label in Figure \ref{pvpos}.  The contour levels are 0.01, 0.02, 0.03, 0.04, 0.05, 0.06, 0.07, 0.08, 0.1, 0.125, 0.15, 0.2, 0.25, 0.3, 0.35, 0.4, 0.45, 0.5, 0.55, and 0.6 Jy beam$^{-1}$. }
\end{figure}

\begin{figure}
\figurenum{13}
\epsscale{0.75}
\plotone{McGary.fig13.eps}
\caption{Position-velocity diagrams for NH\3(2,2) emission.  Position 0 corresponds to the letter label in Figure \ref{pvpos}.  The contour levels are 0.01, 0.02, 0.03, 0.04, 0.05, 0.06, 0.07, 0.08, 0.1, 0.125, 0.15, 0.2, 0.25, 0.3, 0.35, 0.4, 0.45, 0.5, 0.55, and 0.6 Jy beam$^{-1}$. }
\end{figure}

\begin{figure}
\figurenum{14}
\epsscale{0.75}
\plotone{McGary.fig14.eps}
\caption{Position-velocity diagrams for NH\3(3,3) emission.  Position 0 corresponds to the letter label in Figure \ref{pvpos}.  The contour levels are 0.01, 0.025, 0.04, 0.055, 0.075, 0.1, 0.125, 0.15, 0.2, 0.25, 0.3, 0.35, 0.4, 0.45, 0.5, 0.55, and 0.6 Jy beam$^{-1}$. }
\end{figure}

\end{document}